\documentclass[prd,aps,preprintnumbers,floats,floatfix,superscriptaddress,preprintnumbers,showpacs,eqsecnum,nofootinbib,notitlepage]{revtex4-1}

\usepackage[utf8]{inputenc}
\usepackage{latexsym,array,theorem,mathrsfs,bm,float}
\usepackage{psfrag}
\usepackage{amsfonts,amsmath,amssymb,latexsym,array,afterpage,
theorem,mathrsfs,bm,float,epsfig,color,graphicx,tabularx,here,multirow,braket}
\usepackage[colorlinks=false,hidelinks]{hyperref}

\usepackage{comment}

\newcommand*{\ba}{\begin{eqnarray}}
\newcommand*{\ea}{\end{eqnarray}}
\newcommand*{\bb}{\begin{framed}}
\newcommand*{\eb}{\end{framed}}

\newcommand*{\mpl}{M_{\rm Pl}}
\newcommand{\simgt}{\lower.5ex\hbox{$\; \buildrel > \over \sim \;$}}
\newcommand{\simlt}{\lower.5ex\hbox{$\; \buildrel < \over \sim \;$}}

\newcommand*{\Gdot}{{\dot G}}

\newcommand*{\p}{\partial}

\newcommand*{\calA}{{\cal A}}

\newcommand*{\calC}{{\cal C}}
\newcommand*{\calD}{{\cal D}}

\newcommand*{\calE}{{\cal E}}
\newcommand*{\calM}{{\cal M}}

\def\({\biggl(}
\def\){\biggr)}
\def\[{\biggl[}
\def\]{\biggr]}

\begin{document}
\preprint{KUNS-2911}
\title{Linear growth of structure in projected massive gravity}

\author{Yusuke Manita}
\email{manita@tap.scphys.kyoto-u.ac.jp}
\affiliation{Department of Physics, Kyoto University, Kyoto 606-8502, Japan}

\author{Rampei Kimura}
\email{rampei@aoni.waseda.jp}
\affiliation{Waseda Institute for Advanced Study, Waseda University,
	1-6-1 Nishi-Waseda, Shinjuku, Tokyo 169-8050, Japan}
	
\date{\today}

\begin{abstract}
In the present paper, we investigate the linear growth of matter fluctuations based on a concrete model of the projected massive gravity, which is free of the Boulware-Deser ghost and preserves the global Lorentz symmetry. We found that at subhorizon scales, the modification to the linear growth is strongly suppressed even without nonlinear screening of an additional force. In addition, we obtain observational constraints from distance and redshift space distortion measurements and find that there is a parameter region that is consistent both observationally and theoretically.
\end{abstract}

\maketitle

\section{Introduction}
Various cosmological observations such as type Ia supernovae and large-scale structures strongly suggest that  our current Universe is undergoing accelerated expansion \cite{SupernovaSearchTeam:1998fmf,SupernovaCosmologyProject:1998vns,Scolnic:2017caz}\,. However, it is still unclear what causes the accelerated expansion. An attractive approach to this problem is to modify the law of gravity at a long distance. 
Among various modified gravity theories, a relativistic theory with nonvanishing graviton's mass, i.e., massive gravity, is one of the natural and interesting extensions of general relativity, which could potentially explain the accelerated expansion of our Universe.

The construction of a graviton's mass
 originated in the context of a linear theory for a spin-2 particle by Fierz and Pauli in 1939~\cite{Fierz:1939ix}\,. After almost seven decades,  a theoretically consistent theory with a nonlinear mass term was first realized by de Rham, Gabadadze and Tolley~\cite{deRham:2010ik,deRham:2010kj}\,, and it has been called the dRGT theory after their initial letters. Thanks to the special structure of the dRGT potential, it is free from the Boulware-Deser (BD) ghost \cite{PhysRevD.6.3368}, which generally appears at the nonlinear level as a consequence of the breaking of general covariance \cite{Hassan:2011hr}.
Under the assumption of flat Friedmann-Lema\^{i}tre-Robertson-Walker (FLRW), the dRGT theory leads to a trivial solution with a constant scale factor \cite{DAmico:2011eto}. On the other hand, under the assumption of open FLRW, the dRGT theory was shown to have a self-accelerating solution~\cite{Gumrukcuoglu:2011ew}. Unfortunately, the linear perturbation of this solution suffers from a strong coupling problem~\cite{Gumrukcuoglu:2011ew}\,. Furthermore, it was found that the self-accelerating solution of the dRGT theory has a nonlinear instability~\cite{DeFelice:2012mx}\,. Therefore, the dRGT theory does not have a theoretically consistent FLRW solution.

In typical massive gravity, the general covariance is broken due to the presence of a mass term which consists of the physical metric $g_{\mu\nu}$ and a Minkowski metric $\eta_{\mu\nu}$, but it can be restored by introducing the St\"{u}ckelberg fields $\phi^a$ via \cite{Arkani-Hamed:2002bjr}
\ba
f_{\mu\nu}= \eta_{ab} \partial_\mu \phi^a \partial_\nu \phi^b \,,
\ea
where $ \eta_{ab}$ is a Minkowski metric and $a, b = 0, 1, 2, 3$. 
The dRGT theory implicitly assumes inner global Poincar\'{e} symmetry for St\"{u}ckelberg fields, but this assumption can be relaxed for general massive gravity theories. 
 A previous study~\cite{Gumrukcuoglu:2020utx} have shown that a massive gravity with five degrees of freedom, which breaks the inner global translational symmetry but preserves inner global Lorentz symmetry, can potentially have (i) a nonminimal coupling between the scalar curvature and the St\"{u}ckelberg fields and (ii) two classes of mass terms. The first mass term is a straightforward generalization of the dRGT theory, where the constant parameters in the mass potential are promoted to be a function of the St\"{u}ckelberg fields~\cite{DeRham:2014wnv}\,, 
 and it also admits disformal deformation of a field space metric. 
 This theory is an extension of generalized massive gravity (GMG).
 The second mass term is quite different from the dRGT theory, with a field space metric that is the same as the projected tensor on the space perpendicular to the St\"{u}ckelberg field. Then, the theory with this mass term is called projected massive gravity (PMG). 

According to the previous studies~\cite{Kenna-Allison:2019tbu,Gumrukcuoglu:2020utx}, the 
extended theory of GMG and PMG have a self-accelerating solution without theoretical instabilities unlike the dGRT theory. In the case of GMG, the linear growth of density perturbations are investigated in \cite{Kenna-Allison:2020egn}, and the presence of a Vainshtein screening mechanism \cite{Vainshtein1972} has been confirmed in \cite{Gumrukcuoglu:2021gua}. On the other hand, the linear growth of density perturbations as well as background analysis in PMG are still less well understood.

In this paper, we first re-investigate self-accelerating solutions in a concrete model without assuming the weak $X$-dependence and their perturbative stabilities based on the linear theory. Then, we numerically solve evolution equations for matter fluctuation and the scalar mode of a massive graviton, and compare results with one in the quasistatic limit. We further put observational constraints on the model parameter and cosmological parameters from type Ia supernova (SN) and redshift space distortion (RSD) data.

The paper is organized as follows. In Section~\ref{sec:projected_massiv_gravtiy}, we briefly review 
the PMG theory and derive the covariant equations of motion.
In Section~\ref{sec:background_and_linear}, we derive the background equations in a homogeneous and isotropic Universe as well as linear perturbation equations in unitary gauge. In Section~\ref{sec:minimal_coupling_model}, we investigate background dynamics, its stabilities, and the linear evolution of fluctuations in a minimal coupling model. In Section~\ref{secV}, we provide the current constraints from Type Ia SN and RSD data. Section~\ref{sec:conclusion} is devoted to the conclusion. Throughout this paper, we adopt the units with $c=1$ and the notation $\mpl^2=(8\pi G)^{-1}$.

\section{Projected massive gravity}
\label{sec:projected_massiv_gravtiy}

In this section we briefly review the projected massive gravity. The action is given by~\cite{Gumrukcuoglu:2020utx}
\ba
S
&=&
\int {\rm d}^4 x \sqrt{-g} \frac{\mpl^2}{2}\Biggl[ G(X) \,R  - {6 (G'(X))^2 \over G(X)} [Y] 
+ m^2 U(X, [Z], [Z^2], [Z^3])\Biggr]+S_{\rm m}[g_{\mu\nu}, \psi]
\,,\label{action}
\ea
where $S_{\rm m}$ is the matter Lagrangian and
\ba
&&X := { \phi}^a { \phi}_a  \,, \qquad
Z^\mu_{~\nu} := (g^{-1}{\bar f})^\mu_{~\nu} \,,\qquad
Y^\mu_{~\nu} :=  (g^{-1}{\tilde f})^\mu_{~\nu}\\
&&
{\bar f}_{\mu\nu} := P_{ab}\,\partial_\mu { \phi}^a\,\partial_\nu { \phi}^b\,, \qquad 
{\tilde f}_{\mu\nu} := -\phi_a \phi_b\,\partial_\mu { \phi}^a\,\partial_\nu { \phi}^b\,, 
\label{eq:defWY}
\ea
with the projection operator to the field space perpendicular to $\phi^a$,
\ba
    P_{ab} := \eta_{ab} - {{ \phi}_a{ \phi}_b \over X}
    \,.\label{Pab}
\ea
Here, we have used the notation $G'(X)=\frac{\partial G}{\partial X}$ and $[Z^n]$ denotes the trace of matrices $(Z^n)^{\mu}_{~\nu} = Z^{\mu}_{~\alpha_1}Z^{\alpha_1}_{~\alpha_2}\cdots Z^{\alpha_n}_{~\nu}$. 

Due to the presence of $X$-dependent function in the action \eqref{action}, it breaks the translation invariance  $\phi^a \to \phi^a + {\rm const.}$, while the global Lorentz symmetry is preserved. 
The first term in \eqref{action} is the nonminimal coupling between the Ricci scalar $R$ and an arbitrary function of a Lorentz-invariant scalar $X$, and the second term is responsible for eliminating the BD ghost appearing in the first term. The potential $U$ is an arbitrary function of $X$ and $[Z^n]$ and thus there is no need to tune the form of the potential, unlike the dRGT theory.
This is because the projection tensor \eqref{Pab} prevents the BD ghost from appearing in a full theory,
which can be easily shown in the Hamiltonian analysis \cite{Gumrukcuoglu:2020utx}. 

Varying the action with respect to $g_{\mu\nu}$ gives the modified Einstein equation, 
\ba
\mpl^2\left[G(X) G_{\mu\nu}-\nabla_\mu \nabla_\nu G(X) + g_{\mu\nu} \square G - {6 (G'(X))^2 \over G(X)} \left({\tilde f}_{\mu\nu}-{1\over 2}[Y]g_{\mu\nu}\right) \right] = T_{\mu\nu}^{\rm (mass)} +T^{(\rm m)}_{\mu\nu}
\,,\label{eq:eom}
\ea
where $T_{\mu\nu}$ is the energy-momentum tensor for matter content, and $T_{\mu\nu}^{\rm (mass)}$ is the effective energy-momentum tensor of the mass term defined as
\ba
T_{\mu\nu}^{\rm (mass)} =
\mpl^2 m^2 \left(
{1 \over 2} g_{\mu\nu} U - U_{[Z]} {\bar f}_{\mu\nu} 
-2 U_{[Z^2]} Z^{\rho}_{~(\mu} {\bar f}_{\nu)\rho} 
-3 U_{[Z^3]} Z^{\rho}_{~\sigma} Z^{\sigma}_{~(\mu} {\bar f}_{\nu)\rho} 
\right)\,, 
\ea
where we defined $U_{[Z^n]}:=\partial U/\partial [Z^n]$.
We assume that the matter field obeys the standard conservation law,
\ba
\nabla^\mu T^{(\rm m)}_{\mu\nu} =0
\,.\label{eq:conservationlaw}
\ea
Then, the contracted Bianchi identity provides the St\"{u}ckelberg equation,
\ba
\nabla^\mu \left({2\over \sqrt{-g}}{\delta S \over \delta g^{\mu\nu}}\right) = {1 \over \sqrt{-g}} {\delta S \over \delta \phi^a} \partial_\nu \phi^a \,.
\label{StuckelbergEquation}
\ea 
Note that the St\"{u}ckelberg equation is not independent with \eqref{eq:eom} and \eqref{eq:conservationlaw}.

\section{Background equation and linear equation}
\label{sec:background_and_linear}

In this section, we first review background equations with an ansatz respecting FLRW symmetry. We then derive linear perturbation equations in the unitary gauge and derive the master equations and sound speeds for the scalar graviton and the matter fluid. We also find the evolution equation for matter overdensity based on the quasistatic approximation.

\subsection{Background equations}
In order to obtain a homogeneous and isotropic solution, we require FLRW symmetries for both physical metric and fiducial metrics as well as the homogeneity of the Lorentz-invariant scalar $X = \eta_{ab}\phi^a\phi^b$. 
This restricts the line element for the physical metric $g_{\mu\nu}$ to be
an open $(\kappa>0)$ FLRW Universe, 
\ba
    ds_g^2 = -dt^2 + a(t)^2 \Omega_{ij}dx^i dx^j
    \,,
    \label{metric}
\ea
where $\Omega_{ij}$ is an induced metric on the constant time hypersurface defined as
\ba
	\Omega_{ij} = \delta_{ij}-\frac{\kappa\,x^ix^j}{1+\kappa\,x^kx^k}
	\,.
\ea
Then the unique configuration of the St\"{u}ckelberg fields preserving homogeneity and isotropy is given by
\ba
	{ \phi}^0 = f(t)\,\sqrt{1+\kappa(x^2+y^2+z^2)}\,,
	\qquad
	{\phi}^i = f(t)\,\sqrt{\kappa}\,x^i
	\,,\label{eq:Stuckelberg}
\ea
and  correspondingly the fiducial metrics \eqref{eq:defWY} are 
\begin{align}
    ds^2_{\bar{f}}=\bar{f}_{\mu\nu}dx^\mu dx^\nu&=\kappa f^2\Omega_{ij}dx^idx^j \,,\\
    ds^2_{\tilde{f}}=\tilde{f}_{\mu\nu}dx^\mu dx^\nu&={-}f^2\dot{f}^2dt^2 \,.
\end{align}
For matter content, we assume a dust fluid, i.e., the perfect fluid with no pressure,
\ba
     T^{\rm (m)}{}^\mu{}_\nu=\rho u^\mu u_\nu\,,
\ea
where $\rho$ is the energy density of the fluid and $u^\mu$ is a four-velocity.

The Einstein equation \eqref{eq:eom} and the St\"{u}ckelberg equation \eqref{StuckelbergEquation} gives
\ba
3G \left[\left(H+ {\Gdot \over 2G}\right)^2 - {\kappa \over a^2}\right]&=&
{\rho \over \mpl^2}+  {\rho_g\over \mpl^2} \,, \label{eq:Freidmann}\\
-2 G\left[
\p_t \left(H+ {\Gdot \over 2G}\right) + {\kappa \over a^2}
\right]
+ \Gdot \left( H+ {\Gdot \over 2G}\right) 
&=& {\rho \over \mpl^2}+ {\rho_g + p_g \over \mpl^2} \,,
\label{eq:Freidmann2}\\
{\dot \rho}_g + 3H (\rho_g+p_g)-{{\dot G}\over 2G} (\rho_g -3p_g + \rho ) &=&0\,,
\label{StukelbergEqn}
\ea
where the effective energy density $\rho_g$ and pressure $p_g$ for the graviton's mass are defined as
\ba
    \rho_g &=& -{1\over 2}m^2\mpl^2 U
    \,, \label{eq:rhog}
    \\
    p_g &=&{1\over 2}m^2\mpl^2 \Big(U- 2 \xi^2 U_{[Z]}-4\xi^4 U_{[Z^2]}-6\xi^6 U_{[Z^3]}\Big)
    \,,\label{eq:pg}
\ea
and we have also defined
\ba
    \xi = \frac{\sqrt{\kappa}f}{a} \,.
\ea

The conservation law of matter \eqref{eq:conservationlaw} gives the background matter equation,
\begin{align}
    \dot{\rho}+3H\rho=0\,.
    \label{eqmat}
\end{align}
Note that \eqref{eq:Freidmann}, \eqref{eq:Freidmann2}, \eqref{StukelbergEqn}, and \eqref{eqmat} are not all independent, but three of them are independent.

\subsection{Linear perturbations}
In the present paper, we focus on the scalar perturbations to track the evolution of the matter overdensity. For simplicity, we adopt the unitary gauge where perturbations of the St\"{u}ckelberg fields vanish, $\delta\phi^a=0$. In this gauge, the physical metric is defined as
\ba
    ds^2 = g_{\mu\nu} dx^\mu dx^\nu
    = -(1+2\phi) dt^2 + 2a \partial_i B dt dx^i + a^2 (\Omega_{ij} + h_{ij}) dx^i dx^j 
    \,,\label{eq:linearlineelement}
\ea
where 
\ba
    h_{ij} = 2 \psi \Omega_{ij} + \left(D_i D_j - {1\over 3} \Omega_{ij} D_l D^l\right)S
    \,.
\ea
We have defined the covariant derivative $D_i$ on a constant time hypersurface which is associated with the induced metric $\Omega_{ij}$. For matter perturbation, we introduce $\rho(t,\bm{x})=\rho(t)+\delta\rho(t,\bm{x})$ and $u^\mu=(1-\phi,\partial^i v)$. In this set up, the energy-momentum tensor for matter content is given by
\ba
    T^{0}_{~0} &=& -(\rho + \delta\rho)\,,\\
    T^{0}_{~i} &=& a \rho (\partial_i B + a \partial_i v)\,,\\
    T^{i}_{~0} &=& -\rho \partial^i v\,,\\
    T^{i}_{~j} &=& 0\,,
\ea
where $v$ is the 3-velocity potential for matter field. From these perturbations, we define gauge invariant variables~\cite{Bardeen:1980kt}\,,
\ba
    \Phi &:=& \phi - {\dot \zeta}
    \,,
    \\ 
    \Psi &:=& \psi - H \zeta -{1\over 6} D^i D_i S
    \,,
    \\ 
    \tilde{\delta\rho} &:=& \delta\rho - {\dot \rho}\zeta
    \,,
    \\ 
    {\tilde v} &:=& v + {1\over 2} {\dot S}
    \,,
\ea
where
\ba
    \zeta := -a B + {1\over 2} a^2 {\dot S}
    \,.
\ea
We also define a gauge invariant density fluctuation of matter,
\ba
    \Delta := {\tilde{\delta\rho} \over \rho} -3 a^2 H {\tilde v}
    \,.
\ea
Then, we expand all perturbations in scalar harmonics, for example, 
\ba
    \Psi = \int dk k^2 \Psi_{|{\vec k}|} Y({\vec k}, {\vec x})
    \,.\label{eq:psiperturbation}
\ea
Substituting \eqref{eq:linearlineelement}-\eqref{eq:psiperturbation} into the equations of motion \eqref{eq:eom}, we obtain the perturbative equations of motion at the linear level. We show the full expressions in Appendix \ref{sec:linearequations}. 
Note that it is constructed from six perturbations $\{\Phi_{|\vec{k}|},\Psi_{|\vec{k}|},E_{|\vec{k}|},S_{|\vec{k}|},\Delta_{|\vec{k}|},\tilde{v}_{|\vec{k}|}\}$, but only two of them are dynamical. 
As we show in Section \ref{sec:MasterequationWithG1}, we can reduce these equations to two master equations by integrating out the nondynamical variables. 

In the following, we focus on a minimal coupling model where $G=1$, and nonminimal cases are summarized in the Appendix.~\ref{sec:nonminimal_coupling_model}.

\subsection{Master equations with $G=1$}
\label{sec:MasterequationWithG1}
For later convenience, let us derive the master equations for the scalar graviton and  matter perturbation. Although we have six perturbations $\{\Phi, \Psi ,E ,S ,\Delta ,\tilde{v}\}$,  four of them are nondynamical, and we should be able for six linear equations \eqref{Eq:E00p}-\eqref{Eq:Eeup} to reduce to two dynamical equations of motion. To this end, we first solve the continuity equation \eqref{Eq:Ecop}, Euler equations \eqref{Eq:Eeup}, the $(0,0)$ component of the Einstein equation \eqref{Eq:E00p}, and its time derivative in terms of $\tilde v$, $\dot{\tilde v}$, $\Phi$, $\dot \Phi$. Substituting these constraints into the $(0,i)$ component of the Einstein equation \eqref{Eq:E0ip}, one obtains $\Psi$ in terms of $S$, $\Delta$, and their time derivatives. Finally, we arrive at the dynamical equation for $S$ from the traceless equation \eqref{Eq:Etlp} by using the above solutions. The evolution equation for matter perturbation can be obtained by substituting \eqref{Eq:Ecop} and its time derivative into \eqref{Eq:Eeup} to eliminate $\tilde{v}$, $\dot{\tilde{v}}$, and other nondynamical variables. Note that the remaining equation \eqref{Eq:Etrp} can be used for determining $B$. After these procedure, the resultant master equations for $S$ and $\Delta$ are given by 
\ba
    C_1 {\ddot S} + C_2{\dot S} + C_3 S + C_4  \Delta &=& 0 \,,
    \label{eq:masterS} \\
    D_1 {\ddot \Delta} + D_2 {\dot \Delta} + D_3 \Delta + D_4 S &=& 0 \,,\label{eq:masterD}
\ea 
where the coefficients are given by
\ba
C_1 &=& 3 (\rho_g + p_g) \,, \\
C_2 &=& {1\over Q}\Big[
3H(\rho_g+p_g) \Big(
4 \mpl^2 (k^2 + 3\kappa)+9a^2 (\rho_g+p_g)
\Big)
+6\mpl^2 (k^2 + 3\kappa){\dot p}_g
\Big]
\,, \\
C_3 &=& {1\over a^2 H Q}\Big[
H\Big(
M_{\rm GW}^2 Q^2 -3 k^2 a^2 (\rho_g+p_g)^2
\Big)
-2\mpl^2 k^2 (k^2 + 3\kappa){\dot p}_g
\Big] \,, \\
C_4 &=& {6 \rho \over H Q}
\Big[
{\dot p}_g - H(\rho_g+p_g)
\Big] \,, \\
D_1 &=& 1\,, \\
D_2 &=&{1\over Q}\Big[
H \Big(
4 \mpl^2 (k^2 + 3\kappa)+9a^2 (\rho_g+p_g)
\Big)
-3a^2{\dot p}_g
\Big] \,, \\
D_3 &=&-{\rho\over 2 \mpl^2 H Q}\Big[
2 \mpl^2 (k^2 + 3\kappa) H+3a^2{\dot p}_g
\Big] \,, \\
D_4 &=&{1 \over2 a^2 H Q}
\Big[
k^2 (k^2 + 3\kappa)a^2 \Big({\dot p}_g - H(\rho_g+p_g) \Big)
\Big] \,, \\
Q&=& 2 \mpl^2 (k^2 + 3\kappa)+ 3a^2 (\rho_g + p_g)\,.
\ea
One can immediately notice that the ghost instability can be avoided if and only if 
\ba
\rho_g + p_g >0 \,,
\label{condition:ghost}
\ea
which agrees with the result obtained from the quadratic action in \cite{Gumrukcuoglu:2020utx}. 
To find the sound speeds for perturbations, let us assume 
$S(t)=e^{i \omega t} S_0$ and $\Delta(t)=e^{i \omega t} \Delta_0$. Then, the system that consists of \eqref{eq:masterS} and \eqref{eq:masterD} admits the solution if the following condition is satisfied,
\ba
	\left|
    \begin{matrix}
	    -\omega^2 C_1+i\omega C_2+C_3 & C_4\\
	    D_4 & -\omega^2D_1+-i\omega D_2+D_3
    \end{matrix}
    \right| 
    = 0
    \,.
\ea
In the subhorizon limit $k \gg a H$, the above equation provides the standard dispersion relation $\omega^2 \simeq c_s^2 k^2$, where the sound speed of the scalar graviton is given by\footnote{This expression is different from one obtained in the previous paper \cite{Gumrukcuoglu:2020utx}. This is because the matter field was assumed to be $k$-essence, which is essentially different from the perfect fluid at perturbation level.}
\ba
    c_s^2 = \frac{2\mpl^2 H M_{\rm GW}^2 -\dot{p}_g}{3H(\rho_g + p_g)}
    \,,\label{eq:soundspeedminimal}
\ea
and one for the matter field is simply given by $c_s^2=0$.

\subsection{Quasistatic limit of the minimal coupling model}
\label{sec:QS}
In this subsection, we investigate the evolution of the matter density perturbation at subhorizon scales $k \gg a H$, where the data coming from galaxy surveys are applicable. In the quasistatic approximation, we assume that the time scale of the evolution of perturbations is roughly the Hubble time, that is, ${\dot \Psi} \sim H \Psi$, and the density perturbation is roughly given by $\Delta \sim {\cal O}(k^2 \Psi)$. Moreover, the rescaled dimensionless scalar perturbations $\tilde S := k^2 S$ is of order ${\cal O}(\Psi) \sim {\cal O}(\Phi) \ll 1$, which will be confirmed in a numerical analysis in the following section. We further assume that the wavelength of perturbations is well inside the sound horizon, $k^{-1} \ll c_s/(aH)$, where $c_s$ is the sound speed of the scalar mode for the massive graviton given in \eqref{eq:soundspeedminimal}. As will be numerically checked later, the sound speed of the scalar mode is always larger than the speed of light in our case, therefore the obtained equations under the quasistatic approximation are always valid at subhorizon scales. 

The master equations, \eqref{eq:masterS} and  \eqref{eq:masterD}, in the quasistatic limit are given by
\ba
    \ddot{{\Delta}}+2H\dot{{\Delta}}-\frac{\rho}{2 \mpl^2}{\Delta}=0
    \,,\label{eq:tDeltaEvolution}
    \\
    k^2 {\tilde S}=\frac{3a^2\rho [H(\rho_g+p_g)-\dot{p}_g]}{\mpl^2(2\mpl^2M_{\rm GW}^2H-\dot{p}_g)}\Delta
    \,,
\ea
respectively. 

Using the constraint equations for $\Psi$ and $\Phi$ obtained in the previous subsection, we obtain the standard Poisson equation,
\ba
    k^2\Psi=-k^2\Phi=\frac{a^2\rho}{2\mpl^2}\Delta
    \,.\label{eq:SlipPoisson}
\ea
Equation~\eqref{eq:tDeltaEvolution} and \eqref{eq:SlipPoisson} coincide with ones in the $\Lambda$CDM model.
As we will see below, the modification due to the graviton's mass appears only via the modified Hubble expansion at subhorizon scales.

\section{Minimal coupling model}
\label{sec:minimal_coupling_model}

In this section, we consider the minimal coupling model, i.e., $G=1$, with the mass potential
\begin{align}
  U(X,[Z],[Z^{2}],[Z^{3}])=(a_{1}+a_2 m^{2}X)[Z]+{b}[Z]^{2}+{c}[Z^{2}] \,,
  \label{model1}
\end{align}
where $a_1,~a_2,~b$, and $c$ are constant parameters. Here, we require ${a_2}\neq 0$ since the mass potential without $X$-dependence leads to the strong coupling of the scalar perturbation~\cite{Gumrukcuoglu:2020utx}\,.

\subsection{Background solution}

For simplicity, we introduce the following dimensionless parameters :
\ba
    m = \mu H_0 
    \,, \quad 
    \kappa = H_0^2 \Omega_\kappa
    \,, \quad
    \rho = \frac{3H_0^2\mpl^2\Omega_m}{a^3}\,.
\ea
Let us first take a look at the St\"{u}ckelberg equation \eqref{StukelbergEqn}, which is given by
\ba
    \xi(\dot{\xi}+H\xi)\left[ a_{1}\Omega_\kappa+2( {(3b+c)}\Omega_\kappa-a_{2}\mu^{2}a^{2})\xi^{2}\right]=0\,.
\ea
Assuming $\xi$ is nonzero, we get its solutions $\xi\propto 1/a$ and
\ba
  \xi^2=\frac{a_1 \Omega_\kappa }{2 a_2 \mu^2 a^2-2 \Omega_\kappa  {(3b+c)}}
  \,.\label{minimalsol_xi}
\ea
In the first solution $\xi\propto 1/a$, $\rho_{g}$ behaves as the sum of spacial curvature and radiation, so we are not interest in this case. 
Since the fiducial metric should be real, we require
$\xi^2$ to be real for all positive $a$. This yields
\ba
    {3b+c}\geq0~\land~ a_2<0~\land~ a_1<0
    \,.\label{eq:condition_xi2}
\ea
By using the solution \eqref{minimalsol_xi}, the effective equation of state for the graviton's mass is given by
\ba
  w=\frac{p_g}{\rho_g}=-\frac{3 \Omega_\kappa  {(3b+c)}-a_2 \mu ^2 a^2}{3 \Omega_\kappa {(3b+c)}-3 a_2 \mu ^2 a^2}
  \,.\label{eos}
\ea

For convenience, let us introduce the present effective equation of state $w_0 := w|_{a=1}$, and we hereafter use $w_0$ instead of $a_2$, that is, 
\ba
	a_2\mu^2=\frac{3(w_0+1)}{3w_0+1}\Omega_\kappa {(3b+c)}
	\,.\label{eqa2}
\ea
By using \eqref{eqa2}, we can rewrite the equation of state \eqref{eos} as
\ba
	w=-\frac{1+3w_0-a^2(1+w_0)}{1+3w_0-3a^2(1+w_0)}
	\,.\label{eq:eosg}
\ea
This indicates that the equation of state $w$ is solely determined by $w_0$ since $w_0$ absorbs all the dependence of $\mu$ and other parameters. We now consider the Friedmann equation at the present time, and this, for example, determines $a_1$ :
\begin{align}
  a_1\mu=-\sqrt{\frac{-16\Omega_g}{1+3 w_0} {(3b+c)}}
  \,,\label{eqa1}
\end{align}
where $\Omega_g$ is the effective energy density of the mass terms at present defined by $\Omega_g:=1-\Omega_{\rm m}-\Omega_\kappa$. Substituting \eqref{eqa2} and \eqref{eqa1} into the Friedmann equation \eqref{eq:Freidmann}, we obtain
\ba
    \left(\frac{H(a)}{H_0}\right)^2=
    {\Omega_{\rm m} \over a^3} +{\Omega_{\kappa} \over a^2}
    +\frac{2\Omega_g}{3a^2(1+w_0)-(1+3w_0)}
    \,.\label{Hubble}
\ea
Thus, in our model, the background evolution is determined by the four parameters $\{\Omega_{\rm m},\Omega_{\kappa},w_0,H_0\}$\,and identical to the dark energy model with the equation of state \eqref{eq:eosg}.

\begin{figure}[t]
	\begin{tabular}{cc}
		\begin{minipage}{0.5\textwidth}
			\begin{center}
				\includegraphics[scale=0.5]{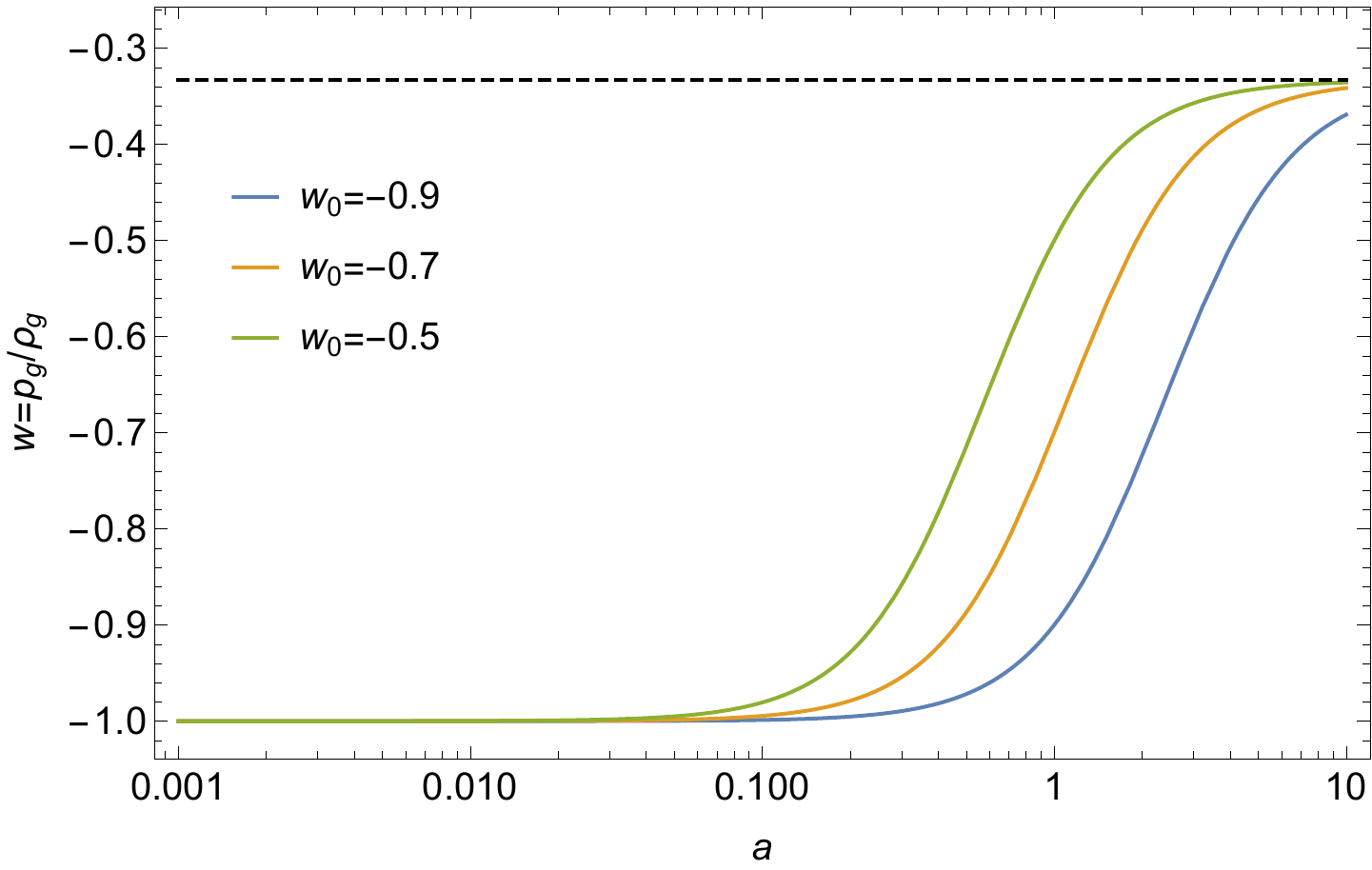}
			\end{center}
		\end{minipage}
		\begin{minipage}{0.5\textwidth}
			\begin{center}
				\includegraphics[scale=0.5]{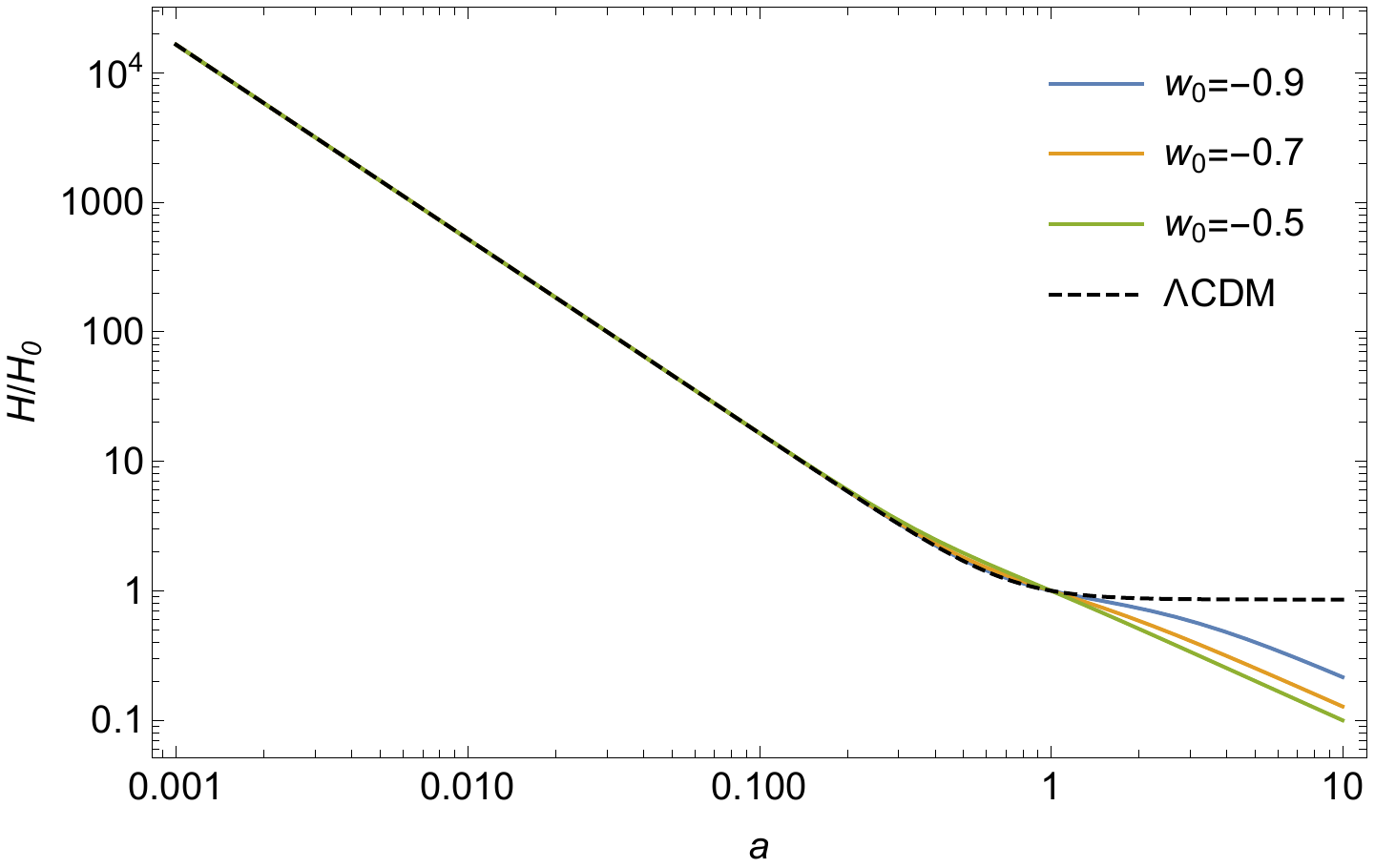}
			\end{center}
		\end{minipage}
	\end{tabular}
	\caption{Left: The evolution of the equation of state for graviton mass term with various $w_0$. Right: The evolution of the Hubble parameter with various $w_0$. {In these figure, we set $\Omega_\kappa=10^{-3}$.} For both plot, solid curves represents PMG with $w_0=-0.9$ (blue), $w_0=-0.7$ (orange) and $w=-0.5$ (green). }
	\label{fig:wandH}
\end{figure}

It is manifest that in the limit of $w_0\to -1$, the equation of state \eqref{eq:eosg} is always unity, and hence the Hubble parameter \eqref{Hubble} reduces to those of $\Lambda$CDM in an open Universe. 
However, this case where $w_0=-1$ corresponding to no $X$-dependence in arbitrary functions, $a_2=0$, leads to the vanishing of the kinetic term for a scalar graviton, which will be discussed in the next subsection.
As can be seen in the left panel of Figure \ref{fig:wandH}, as long as $-1 < w_0 < -1/3$, the equation of state $w$ approaches $-1$ in the past, implying that the graviton mass behaves as a cosmological constant. 
On the other hand, in the limit of $a\to \infty$, the equation of state $w$ approaches $-1/3$. 
At the early stage of the Universe,  the evolution of the Hubble parameter for any $w_0$ is identical to that of the $\Lambda$CDM model as can be checked in the right panel of Figure \ref{fig:wandH}. Around the transition time, the equation of state $w$ starts to increase from $-1$ towards $-1/3$, and the Hubble parameter correspondingly deviates from that of $\Lambda$CDM.

\subsection{Stability conditions}

In this subsection, we derive conditions for avoiding ghost, gradient, and  tachyonic instabilities of the massive graviton on top of \eqref{eq:condition_xi2}. The ghost instability can be avoided if \eqref{condition:ghost} is satisfied. This translates 
\ba
	\frac{\rho_{g}+p_g}{H_0^2\mpl^2}=-\frac{a_1^2 a_2  \mu ^4  \Omega_\kappa  a^2}{4 [\Omega_\kappa  {(3 b+c)}-a_2 \mu^2 a^2]^2} >0 
	\quad
	\Longleftrightarrow\quad a_2 < 0
	\,.
\ea 
In order to avoid the tachyonic instability for tensor and vector perturbations, we impose the positivity of their effective mass squared. According to \cite{Gumrukcuoglu:2020utx}, both tensor and vector mass are given by\footnote{The quadratic Lagrangians for the tensor and vector modes are completely the same as the previous result~\cite{Gumrukcuoglu:2020utx} with $G=1$ because we consider the dust fluid, and it does not contribute to either tenor mode or vector mode. } 
\ba
    M_{\rm GW}^2 =
    {2 \over \mpl^2} \Big[
    \rho_g + p_g - 2m^2 \mpl^2 \xi^4 
    \left(
        U_{[Z^2]}+3\xi^2 U_{[Z^3]} 
    \right)
    \Big]\,.
\ea
Using the solution \eqref{minimalsol_xi}, the condition for avoiding the tachyonic instability is given by
\ba
	\frac{M_\mathrm{GW}^2}{H_0^2} =
	-\frac{a_1^2\mu^2\Omega_\kappa(2{c}\Omega_\kappa+ a_2\mu^2 a^2)}{2\left[{(3 b+c)}\Omega_\kappa -a_2\mu^2 a^2\right]^2}>0
	\,,\quad
	\Longleftrightarrow\quad {c}<0
	\,.
\ea
To avoid gradient instability, we impose that the sound speed for the scalar graviton \eqref{eq:soundspeedminimal} is positive, that is,
\ba
  c_{s}^{2} =
  \frac{\mathcal{G}(a^2)}{3  a_2 \mu^2 a^2 \left[ a_2 \mu^2 a^2 -{(3 b+c)}\Omega_\kappa \right]} >0
  \,,\label{minimal_sound_speed}
\ea
where
\ba
    \mathcal{G}(a^2)
    :=3 a_{2}^2 \mu^4 a^4+3 a_2({b+3c}) \Omega_\kappa \mu^{2}a^{2}-8 {c(3b+c)}\Omega_\kappa^{2}
    \,.
\ea
Since \eqref{eq:condition_xi2} ensures that the denominator of \eqref{minimal_sound_speed} is positive for any $a>0$, we need to impose $\mathcal{G}(a^2) \geq 0$. We can regard $\mathcal{G}(a^2)$ as a quadratic equation for $a^2$, and its discriminant is
\ba
    \calD={3 a_2^2 (3b^2+114b c+59c^2)}\mu^4 \Omega_\kappa^2\,.
\ea
In the case of $\calD\leq0$, $\mathcal{G}(a^2)$ is clearly non-negative for all $a>0$. On the other hand, in the case of $\calD>0$, the larger solution of $\mathcal{G}(a^2)=0$ should be negative:
\ba
    {\frac{\Omega_\kappa}{6a_2\mu^2}\left[-3(b+3c)+\sqrt{3(3b^2+114b c+59c^2)}\right]}
    < 0 
    \,.\label{Dpositive}
\ea
If $b+3c<0$, the condition \eqref{Dpositive} is always satisfied. On the other hand, in the case of $b+3c\geq0$, the condition \eqref{Dpositive} never holds because 
\ba
    {\left[3(b+3c)\right]^2-3(3b^2+114bc+59c^2)=-96c(3b+c)>0}
    \,.
\ea

In summary, the parameter region without ghost, tachyon and gradient instabilities is
\ba
    {a_1<0~\land~ a_2<0 ~\land~ c<0 ~\land~ -\frac{1}{3}\leq {\frac{b}{c}} \leq \frac{59}{-57+32\sqrt{3}}}
    \,.\label{minimal_condition}
\ea
From \eqref{minimal_condition} and \eqref{eqa2}, the stability condition in terms of $w_0$ is
\ba
    -1 < w_0 <-\frac{1}{3}
    \,.\label{eq:stableEOS}
\ea

\subsection{Numerical solutions}

Let us now solve the perturbed equations \eqref{eq:masterS} and \eqref{eq:masterD} numerically. Although the background evolution is completely determined by the effective equation of state at present $w_0$, the perturbation equation now depends on $b$ and $c$ due to the presence of $M_\mathrm{GW}^2$. In the present paper, we, for simplicity, fix these free parameters as 
\ba
	{b}=1
	\,,\quad
	{c}=-1
	\,,\label{b1c1}
\ea
which satisfies the conditions \eqref{minimal_condition} as long as $-1< w_0 < -1/3$. For other cosmological parameters, we adopt  $\Omega_\kappa = 10^{-3}$ and $\Omega_{\rm m} = 0.3$ in this subsection. For a numerical computation, we introduce the dimensionless perturbation,
\ba
    {\tilde S} := k^2 S \,.
\ea

As discussed in the previous subsection, at early times, the equation of state $w$ is close to $-1$, thus the background evolution is almost the same as one in $\Lambda$CDM in an open Universe. For this reason, for the initial conditions of the density perturbations, we adopt the GR solution during matter dominated era :
$\Delta_{\rm ini} = a_{\rm ini}$ and ${\dot \Delta}_{\rm ini} =1 $, where $a_{\rm ini} $ is the initial scale factor for numerical calculation. As for the initial conditions for the scalar graviton, we fix as ${\tilde S}_{\rm ini}  =10^{-10}$ and ${\dot {\tilde S}}_{\rm ini}  =0$. Our numerical calculation shows that the late-time behavior of $\Delta$ and $\tilde{S}$ does not depend on the initial condition of ${\tilde S}_{\rm ini}$, 
and the attractor behavior of $\tilde S$ has been thus confirmed numerically.

\begin{figure}[t]
	\begin{tabular}{cc}
		\begin{minipage}{0.5\textwidth}
		\centering
            \includegraphics[scale=0.5]{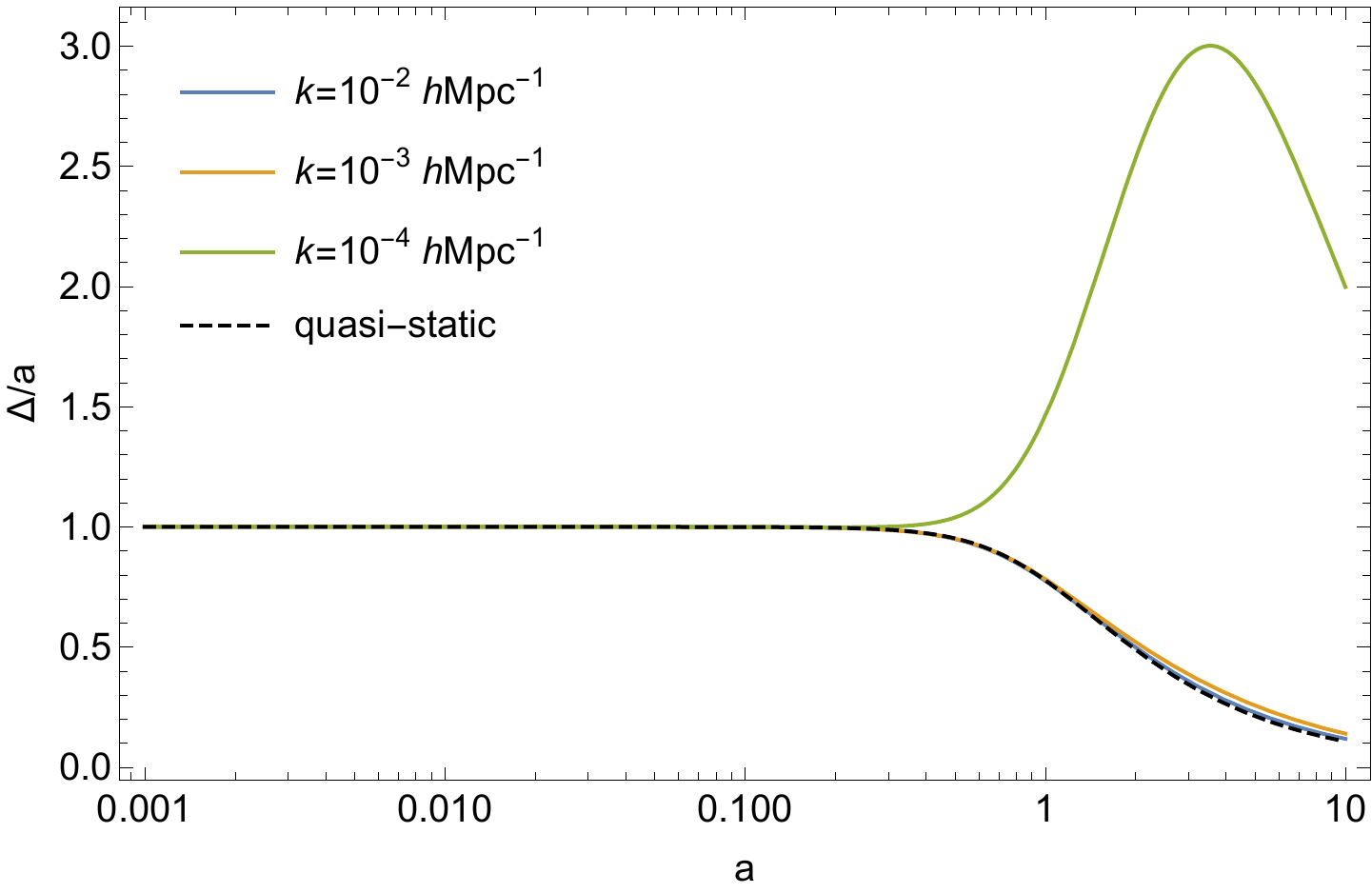}
		\end{minipage}
		\begin{minipage}{0.5\textwidth}
		\centering
            \includegraphics[scale=0.5]{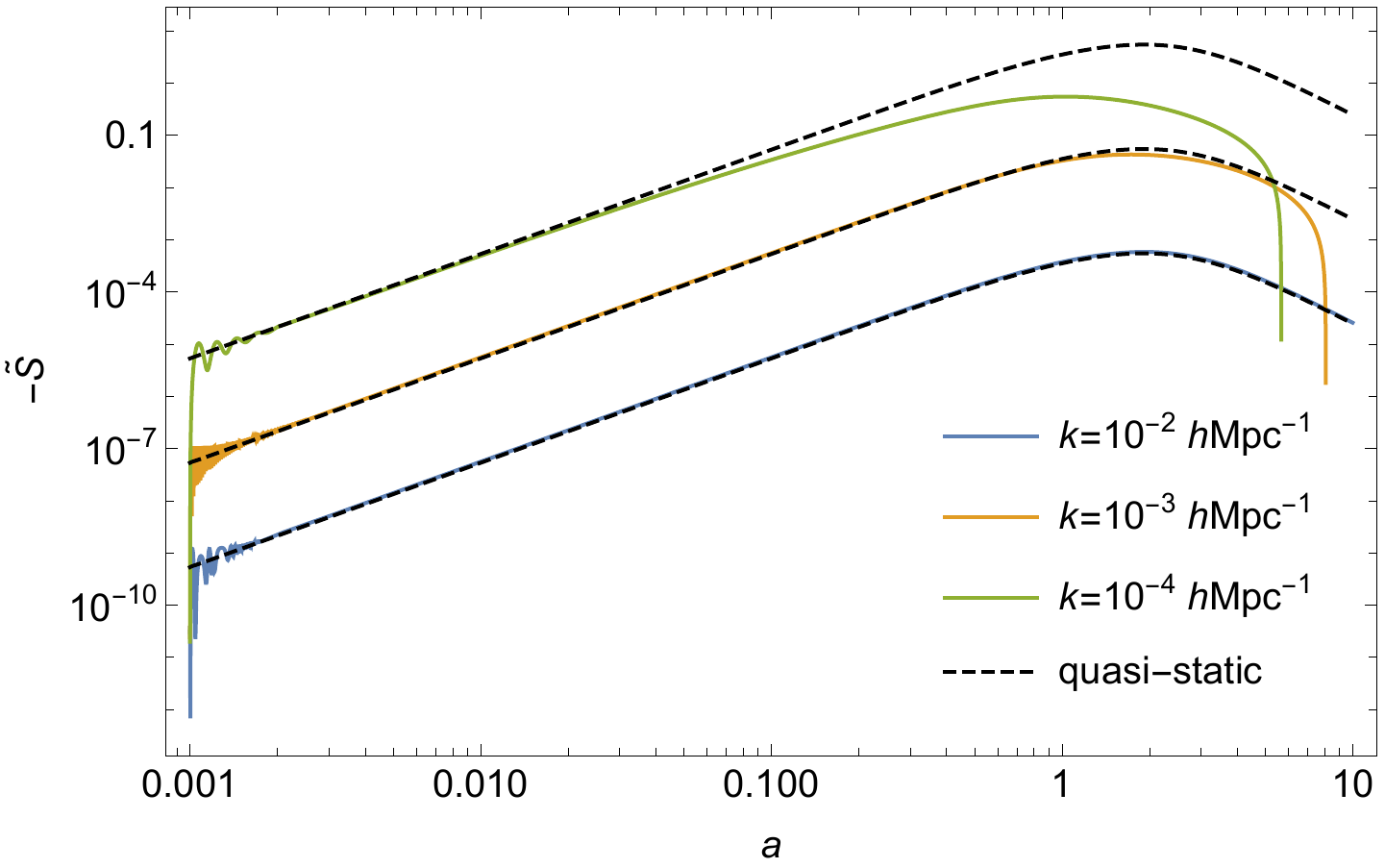}
		\end{minipage}
	\end{tabular}
	\caption{The evolution of the density fluctuation (left) and the graviton scalar perturbation (right) with the wave number $k=10^{-2}h{\rm Mpc}^{-1}~(\mathrm{blue}),10^{-3}h{\rm Mpc}^{-1}~(\mathrm{orange})$ and $10^{-2}h{\rm Mpc}^{-4}~(\mathrm{green})$. We set the current equation of motion as $w_0=-0.9$ for both panels.}
	\label{fig:differentk}
\end{figure}

\begin{figure}[t]
	\begin{tabular}{cc}
		\begin{minipage}{0.5\textwidth}
		\centering
            \includegraphics[scale=0.5]{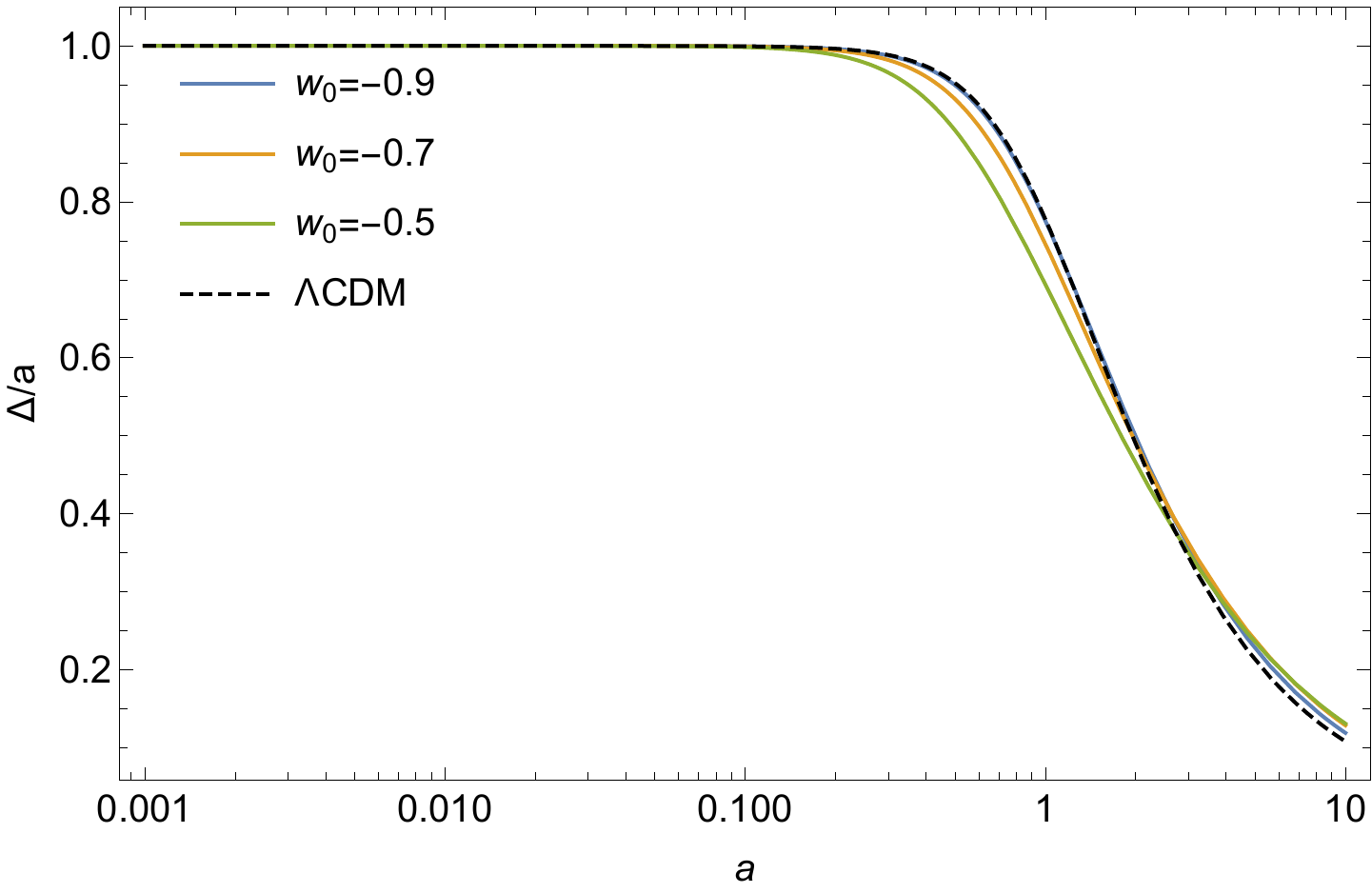}
		\end{minipage}
		\begin{minipage}{0.5\textwidth}
		\centering
            \includegraphics[scale=0.5]{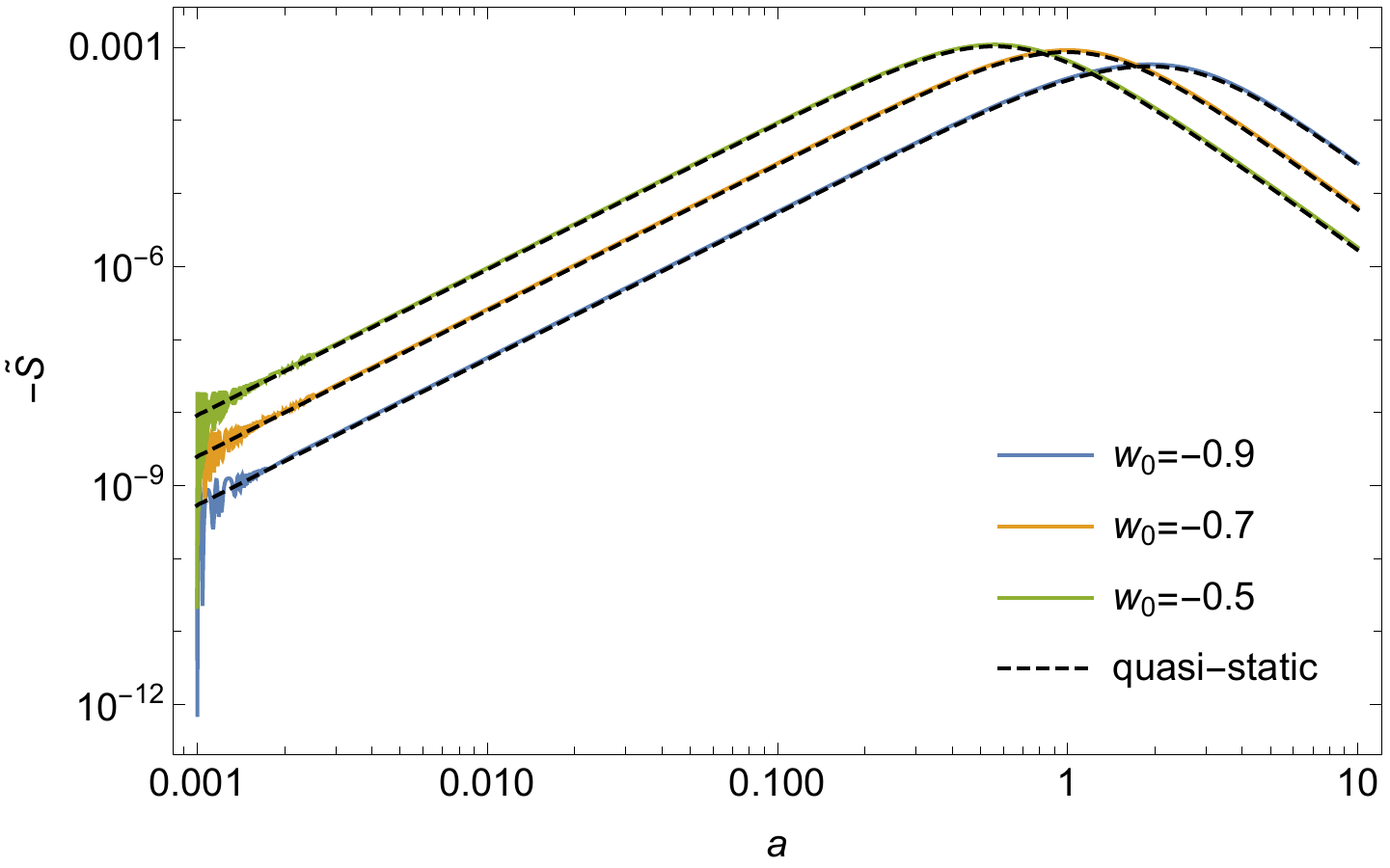}
		\end{minipage}
	\end{tabular}
    \caption{The evolution of the density fluctuation with the current equation of state $w_0=-0.9~(\mathrm{blue}),-0.7~(\mathrm{orange})$ and $-0.5~(\mathrm{green})$. We set the wave number as $k=0.01h{\rm Mpc}^{-1}$ for both panels.}
    \label{fig:differentw0}
\end{figure}

\begin{figure}[t]
    \centering
    \includegraphics[scale=0.5]{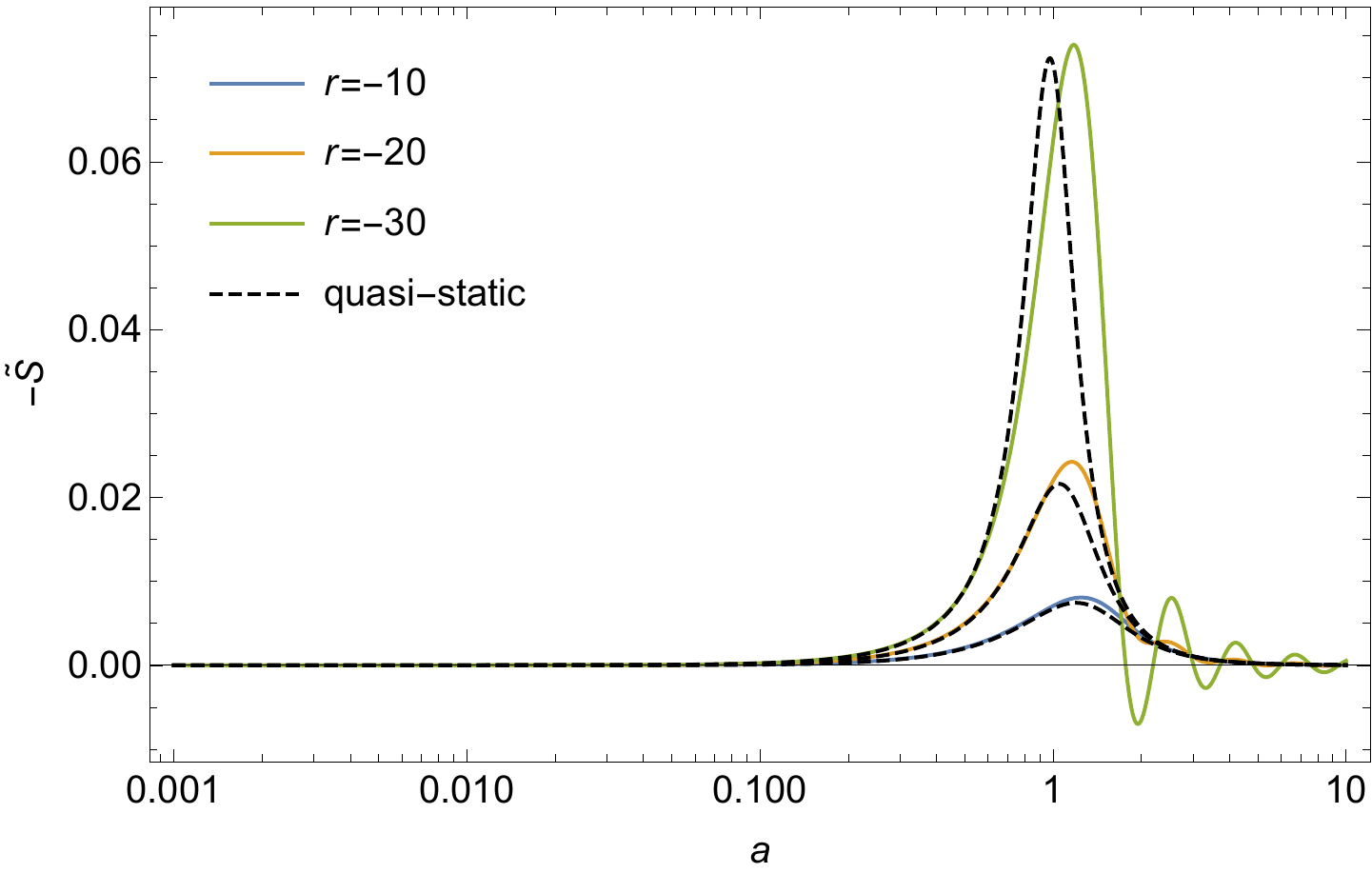}
    \caption{The evolution of the scalar graviton with the parameter $r:=b/c=-10~(\mathrm{blue}),-20~(\mathrm{orange})$ and $-30~(\mathrm{green})$. The black dashed curves represent the corresponding quasistatic limit.
    We set the wave number as $k=0.01h{\rm Mpc}^{-1}$, and the current equation of state $w_0$ as $-0.9$.}
    \label{fig:differentr}
\end{figure}

Figure $\ref{fig:differentk}$ shows the evolution of the normalized density fluctuation $\Delta/a$ (Left panel) and the graviton scalar mode $-\tilde{S}$ (Right panel) with various wave numbers. One can easily confirm that the full numerical calculation agrees with the expectation from the quasistatic limit within subhorizon scales while 
it breaks down when the wavelength is comparable to the horizon scales. 
One should notice that the growth rate at superhorizon scales is strongly enhanced near the present time, and this
is caused by the opposite sign of the Hubble friction term in the master equation \eqref{eq:masterD} only at large scales. For these wavenumbers, the density perturbation exceeds unity at late-time, and our linear perturbation analysis is thus no longer applied. 
Figure~$\ref{fig:differentw0}$ shows the evolution of the normalized density fluctuation $\Delta/a$ with various $w_0$ where the wave number is fixed as $k=10^{-2}h{\rm Mpc}^{-1}$. Since the evolution equation for the density perturbation at subhorizon scales is identical to that of the $\Lambda$CDM model as in \eqref{eq:tDeltaEvolution}, the slight modification to the $\Delta$ is due to the different Hubble expansion, which is only controlled by $w_0$ and irrelevant to the parameters $b$ and $c$.
Finally, let us discuss the evolution of the scalar graviton $\tilde S$. In Fig.~\ref{fig:differentr}, we plot the scalar perturbation $-\tilde{S}$ with various $r := b/c$. The quasistatic results (dashed line) roughly follow the numerical ones at early times, but it slightly deviates at late-time. This is because the sound speed of the scalar graviton becomes small for large $r$ around the present time, implying that even subhorizon perturbations exit the sound horizon of the scalar graviton. Thus, for large $r$, the scalar perturbation $\tilde S$ tends to oscillate as in Fig.~\ref{fig:differentr}. However, this deviation does not significantly affect the evolution of the density perturbation $\Delta$ itself, and the quasistatic approximation is sufficiently valid for  density perturbation at subhorizon scales.

\section{Current constraints}
\label{secV}
In this section, we present observational constraints on the minimal coupling model defined in Section \ref{sec:minimal_coupling_model} by using 1048 SN Ia dataset~\cite{Scolnic:2017caz} and 64 RSD dataset~\cite{Kazantzidis:2018rnb}, which can potentially constrain the model parameter $w_0$ and other cosmological parameters in our case.

\subsection{Type Ia Supernova}

The apparent magnitude of supernovae at redshift $z$ is given by
\ba
    m(z) = \calM+5 \log _{10}\left(H_0 D_L(z)\right)
    \,,\label{eq:def2}
\ea
where $\calM$ is defined by
\ba
    \calM := M+5\log_{10}\left(\frac{1/H_0}{1\mathrm{Mpc}}\right)+25\,,
\ea
with an absolute magnitude $M$. In the case of an open FLRW Universe \eqref{metric},
the luminosity distance $D_L(z)$ is given by
\ba
    D_L(z) := (1+z)\frac{1/H_0}{\sqrt{\Omega_\kappa}}\sinh \left(\frac{\sqrt{\Omega_\kappa}}{1/H_0} D_C(z) \right)\,,
\ea
where $D_C$ is a line-of-sight comoving distance defined by
\ba
    D_C(z) := \frac{1}{H_0}\int_0^z \frac{dz'}{H(z')/H_0}\,.
\ea
In our minimal coupling model of PMG, the normalized Hubble parameter $H(z)/H_0$ can be determined by the three cosmological parameters $\Omega_\kappa, \Omega_{\rm m}$ and $w_0$ as in \eqref{Hubble}, and thus the apparent magnitude $m(z)$ is parametrized by $\{\calM, \Omega_{\rm m},\Omega_\kappa,w_0\}$.

Following \cite{Kazantzidis:2020tko}, we define $\chi_{\rm SN}^2[\calM,\Omega_{\rm m},\Omega_\kappa,w_0]$ by
\begin{align}
    \chi_{\rm SN}^{2}[\calM,\Omega_{\rm m},\Omega_\kappa,w_0]&=\sum_{i=1}^{N}\sum_{j=1}^{N} V_i[\calM,\Omega_{\rm m},\Omega_\kappa,w_0]~\Sigma_{ij}^{-1}~V_i[\calM,\Omega_{\rm m},\Omega_\kappa,w_0]
    \,,\label{eq:chi2_SN}
\end{align}
where
\begin{align}
    V_i[\calM,\Omega_{\rm m},\Omega_\kappa,w_0]:=&m^\mathrm{obs}_i(z_i)-m^\mathrm{th}_i(z_i)=m^{\rm obs}_i(z_i)-\left[\calM+5 \log _{10}\left(\frac{H_0 D_L(z_i)}{1 {\rm Mpc}}\right)+25\right]\,.
\end{align}
We define $N$ as the number of data, and $\Sigma_{ij}$ as a covariant matrix which composed by statistical error and systematic error.

\subsection{Redshift space distortion}

The spatial distribution of galaxies obtained by the RSD survey is different from that in real space. This is because the observed light undergoes a Doppler shift in the line-of-sight direction due to the peculiar velocity of each galaxy. The density fluctuation in real space $\Delta(\bm{k})$ relate to that in redshift space $\Delta^{(\rm S)}(\bm{k})$ by the Kaiser formula \cite{Kaiser:1987qv},
\begin{align}
    \Delta^{({\rm S})}(\bm{k})=(b+f\mu_k^2)\Delta(\bm{k})\,,
\end{align}
where $b$ is a galaxy bias, and $\mu_k$ is a directional cosine of $\bm{k}$ with respect to the line of sight, and $f$ is a linear growth factor defined by
\begin{align}
    f(a):=\frac{d \ln \Delta (a)}{d \ln a}\,.
\end{align}
Through the Kaiser formula, the RSD survey provides the following combination, 
\begin{align}
    f \sigma_{8}(a) := f(a) \cdot \sigma(a)=\frac{\sigma_{8}}{\Delta(1)} \frac{d\Delta(a)}{d \ln a}\,.
\end{align}
The parameter $\sigma_8$ is the normalization factor of the matter power spectrum. In the present paper, we use 63 datasets of the RSD survey listed in \cite{Kazantzidis:2018rnb}. Since each RSD data assumes a different fiducial cosmology, it leads to a false anisotropy in a galaxy power spectrum, which is known as Alcock-Panczy\'{n}ski effect~\cite{Alcock:1979mp}.
Following \cite{Macaulay:2013swa}, we introduce the collection factor $q$ for $f\sigma_8$ to evade the false anisotropy from Alcock-Panczy\'{n}ski effect :
\begin{align}
    f\sigma_8(z)\simeq \frac{H(z)D_A(z)}{H^{\rm fid}(z)D^{\rm fid}_A(z)}f\sigma_8^{\rm obs} := q~f\sigma_8^{\rm obs} \,,
\end{align}
where an angular diameter distance is defined by $D_A(z):=(1+z)^{-2}D_L(z)$, and the sign "fid" represents the value of fiducial cosmology for each observational data.

Following \cite{Kazantzidis:2018rnb}, we define $\chi_{\rm RSD}^2$ by
\begin{align}
    \chi_{\rm RSD}^{2}[w_0,\Omega_{\rm m},\Omega_\kappa,\sigma_8]=\sum_{i=1}^{N}\sum_{j=1}^{N}V_i\left[w_0,\Omega_{\rm m},\Omega_\kappa, \sigma_{8}\right]~\Sigma_{ij}^{-1}~V_i\left[w_0,\Omega_{\rm m},\Omega_\kappa, \sigma_{8}\right]
    \,,\label{eq:chi2_RSD}
\end{align}
where
\begin{align}
    V^{i}\left[w_0, \Omega_{\rm m},\Omega_\kappa, \sigma_{8}\right] := f \sigma^{\rm obs}_{8, i}-q ~f \sigma^{\rm th}_{8}\left[w_0, \Omega_{\rm m}, \Omega_\kappa, \sigma_{8}\right]\,.
\end{align}

\subsection{Observational constraints}
For SN Ia dataset, we use the Pantheon SN Ia sample of $1048$ measurements containing PS1, low-z (\cite{Rest_2014} and reference therein), SDSS \cite{Frieman:2007mr}, SNLS \cite{CGSRA2010,SGCRA2011}, and HST \cite{Riess:2017lxs}, whose range of the redshift is $0.01 < z < 2.3$ ~\cite{Scolnic:2017caz}. The RSD data contains 63 $f \sigma_8$ measurements including e.g., SDSS \cite{Song:2008qt,spr2012,tpbbj2012,Sanchez:2013tga,Howlett:2014opa,Feix:2015dla,Chuang:2013wga,Feix:2016qhh,Shi:2017qpr,Zhao:2018gvb}, BOSS \cite{BOSS:2016wmc,BOSS:2016psr,Gil-Marin:2016wya,Wang:2017wia}, and WiggleZ \cite{Wiggle2012} spanning in the redshift range, $0.001 < z < 1.944$~\cite{Kazantzidis:2018rnb}. 
In our analysis, we define the combined chi square $\chi_{\rm comb}^2$ as a sum of $\chi_{\rm RSD}^2$ and $\chi_{\rm SN}^2$, and the parameters $\cal M$, $\Omega_{\kappa}$ and $\sigma_8$ are marginalized to obtain a two dimensional contour plot.

\begin{figure}[h]
    \centering
    \includegraphics[scale=0.25]{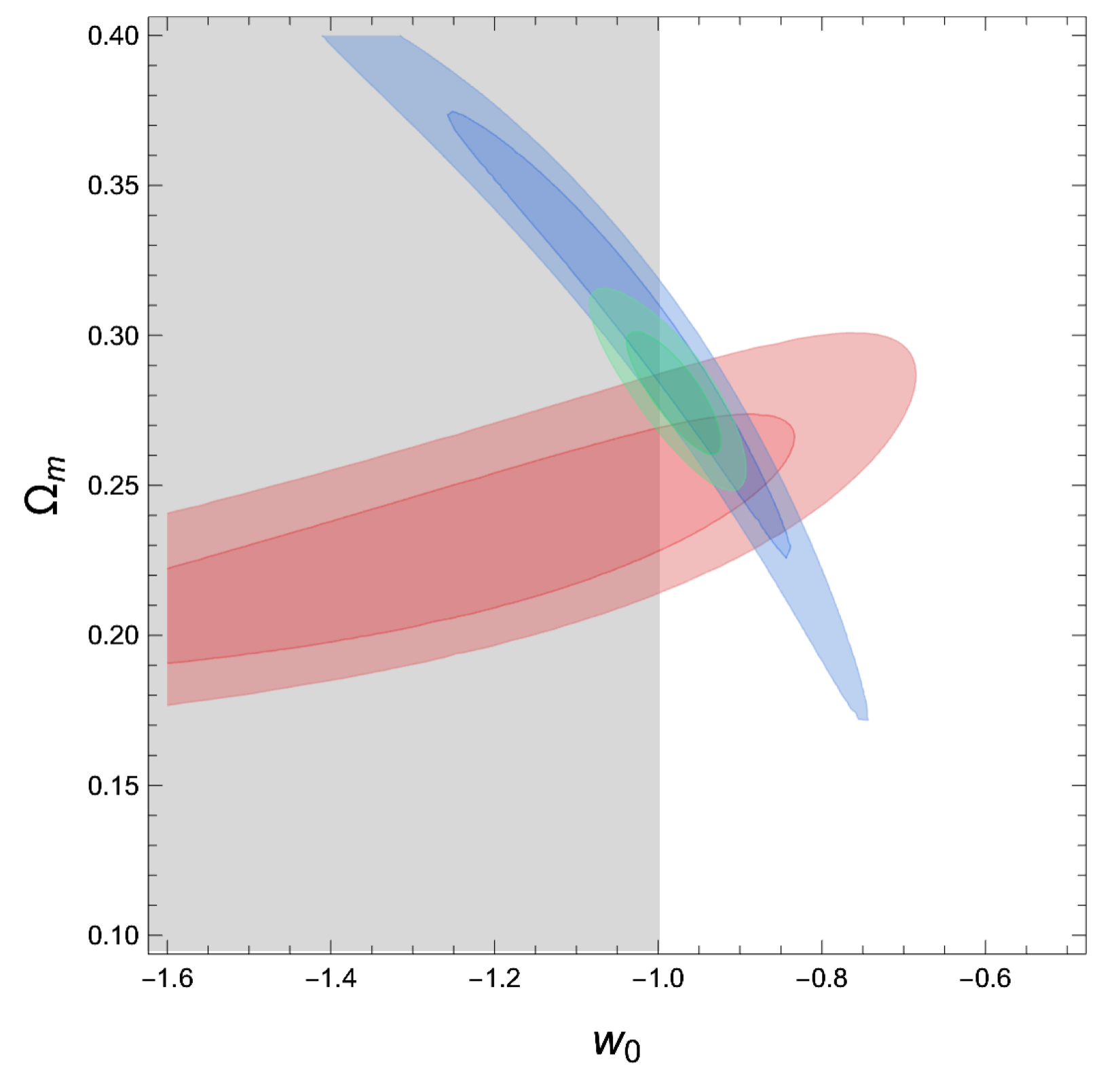}
    \caption{The 68\% and 95\% C.L. contour of $\Omega_{\rm m}$ and $w_0$. The blue contours correspond to the best fit region with the Pantheon dataset \cite{Scolnic:2017caz}. The red contours correspond to the best-fit region with the RSD dataset \cite{Kazantzidis:2018rnb}. The green contour correspond to the dataset best fit region with these combined dataset.}
    \label{fig:NSRSDconstraint}
\end{figure}

Figure \ref{fig:NSRSDconstraint}, we show the 68\% and 95\% C.L. contour of the two-dimensional marginalized constraints in $w_0-\Omega_{\rm m}$ plane. In this figure, the blue, red, and green regions correspond to the constraints from the SN dataset, RSD dataset, and the combined dataset, respectively. The gray shaded region conflicts with the stable condition \eqref{eq:stableEOS}. 
As one can clearly see in Fig. \ref{fig:NSRSDconstraint}, there is a consistent parameter region in $w_0>-1$ as well as the $\Lambda$CDM model corresponding to $w_0=-1$.  Note that neither background and linear growth rate within subhorizon scales is insensitive to the model parameters $b$ and $c$.

\section{Conclusion}
\label{sec:conclusion}
In this paper, we have studied the linear growth of the density fluctuation in the PMG theory, which is the Lorentz-invariant massive gravity with a new graviton mass term.  Investigating the background evolution of the minimal coupling model whose mass term contains up to quadratic order in the trace of $Z$, we found that there exists a self-accelerating solution driven by the graviton mass, without assuming weak $X$-dependence as investigated in \cite{Gumrukcuoglu:2020utx}. The equation of state of the graviton mass approaches $-1/3$ while it behaves as the cosmological constant at an early time. This equation of state can be determined by one parameter $w_0$, which can be described by a combination of the constant parameters appearing in the graviton mass. Then we have derived the linear equations of motion in unitary gauge as summarized in Appendix~\ref{sec:linearequations} and found the stability conditions for avoiding ghost, gradient, and tachyonic instabilities, which can be easily satisfied in our model. The quasi-static approximation, which is valid at subhorizon scales, shows that the evolution equation of the density perturbation is identical to that of the $\Lambda$CDM model, and the modification due to the scalar graviton only appears through the different expansion rate. On the other hand, we numerically confirmed that the matter density fluctuation tends to be large at superhorizon scales. To summarize these results, the background evolution and the linear growth of structure at subhorizon scales in the PMG theory can be regarded as the dark energy model with the equation of state \eqref{eq:eosg}. We finally obtained observational constraints on $\Omega_{\rm m}$ and $w_0$ from SN Ia and RSD dataset, showing that our model is consistent with these observations.

Since our minimal coupling model approaches the $\Lambda$CDM model as $w_0$ approaches $-1$ at both background and linear level and is consistent with $w_0=-1$ as in Fig.~\ref{fig:NSRSDconstraint}, it would be difficult to distinguish our  model from the $\Lambda$CDM model at least at the linear level. Thus, it might be interesting to investigate astrophysical objects, such as black holes and neutron stars, and local gravity tests including whether our model needs a screening mechanism.   

Finally, we have also discussed the nonminimal coupling model in Appendix \ref{sec:nonminimal_coupling_model}, whose arbitrary function is given by $G=(1+g m^2 X)^2$. Taking into account up to a cubic order in the trace of $Z$, we found the stable parameter region in the limit of $a\to 0$ and $a\to \infty$. However, our numerical analysis shows that there exist perturbative instabilities at present even for these parameter regions. This does not conclude that a nonminimal coupling always introduce instabilities, and a further investigation will be necessary in future work.

\begin{acknowledgements}
We would like to thank Masroor C. Pookkillath, Atsushi Taruya, and Daisuke Yamauchi for insightful comments. Y.M. acknowledges the xPand package \cite{Pitrou:2013hga} which was used for confirming calculations explicitly. The work of Y.M. was supported by the establishment of university fellowships towards the creation of science technology innovation.
\end{acknowledgements}

\appendix

\section{Linear perturbation equations}
\label{sec:linearequations}

In this appendix, we summarize the $(0,0)$, $(0,i)$, trace, and traceless component of the Einstein equation, denoted as $\calE^{00}, \calE^{0i}, \calE^{\rm tr}$, and $\calE^{\rm tl}$, and the continuity $\calE^{\rm co}$ and Euler equations $\calE^{\rm eu}$ for the matter fluid.
\ba
    \calE^{00} &=& 
    3G\left(2H + {{\dot G} \over G}\right)^2 \Phi
    -{3\over 2}G\left(2H + {{\dot G} \over G}\right)
    \left(4{\dot\Psi}+2{{\dot G}\over G}a {\dot B} -a^2 {{\dot G}\over G} {\ddot S}\right)
    -\left( {4(k^2 + 3\kappa)\over a^2}G + {6(\rho_g+p_g)\over \mpl^2}\right)\Psi \notag\\
    &&+6a^2 H {\rho \over \mpl^2 }{\tilde v}
    + k^2 {\rho_g+p_g \over \mpl^2} S 
    + {2\rho \over \mpl^2}\Delta +\left((k^2+3\kappa){\dot G} + 3a^2 {\rho + \rho_g+p_g\over 2\mpl^2}{{\dot G}\over G}\right)
    \left({\dot S}-2{B\over a}\right)\notag\\
    &&
    +{3\over 4} a^2 G \left(2H+{{\dot G}\over G}\right)
    \Bigg[
    \left\{
    2\left(H {{\dot G}\over G}+{{\ddot G}\over G}\right)
    -3 {{\dot G}^2\over G^2}
    \right\} {\dot S}
    +2\left(
    3{{\dot G}^2\over G^2}-2{{\ddot G}\over G}
    \right){B \over a}
    \Bigg] \label{Eq:E00p}
    \,,\\
    \calE^{0i} &=& 
    -4G \Bigg[\left(H+{{\dot G}\over 2 G}\right)\Phi- {\dot \Psi}\Bigg]
    -{2a^2 \rho \over \mpl^2}{\tilde v} + 2 a {\dot G} {\dot B} + a \left(2 {\ddot G}-{3 {\dot G}^2 \over G}\right)B
    -a^2 {\dot G} {\ddot S} -a^2 {\rho_g+p_g \over \mpl^2}{\dot S} \notag\\
    &&- a^2 \left({\ddot G} + H {\dot G} -{3{\dot G}^2 \over 2 G}\right) {\dot S}
    \,,\label{Eq:E0ip}\\
    \calE^{\rm tr} &=& 
    6G \left(2H + {{\dot G} \over G}\right) {\dot \Phi}
    -12 G \left(3H + {{\dot G} \over G}\right) {\dot \Psi}
    - \left(4G {k^2-3\kappa \over a^2} + {12 p_g \over \mpl^2} \right)\Phi
    - \left(4G {k^2+3\kappa \over a^2} -6 \calA \right)\Psi
    -12 G {\ddot \Psi} \notag\\
    &&-k^2 \calA S
    + {3a^2 \over 2}G \left(4 {{\ddot G}\over G} + 12H {{\dot G}\over G} -3 {{\dot G}^2 \over G^2}\right) {\ddot S}
    - 3a G \left(4 {{\ddot G}\over G} + 8H {{\dot G}\over G} -3 {{\dot G}^2 \over G^2}\right) {\dot B} \notag\\
    &&-a\left(3\calC_B +4 {\dot G} {k^2 + 3\kappa \over a^2}\right) B 
    + a^2\left(\calC_{\dot S} + 2 {\dot G} {k^2 +3\kappa \over a^2}\right) {\dot S} + 3a {\dot G} (a {\dddot S} -2 {\ddot B})\label{Eq:Etrp}
    \,,\\
    \calE^{\rm tl} &=& -2(\Phi + \Psi) -2 a {{\dot G}\over G}B + {a^2 \over G}(M_{\rm GW}^2 S + {\dot G}{\dot S})\,,\label{Eq:Etlp}\\
    \calE^{\rm co} &=& \Bigg[2(k^2 + 3\kappa)+ {3a^2 \over G} {\rho + \rho_g+p_g \over \mpl^2}
    -3a^2 \left(
    4H^2 + {{\dot G} \over G}H + {3 {\dot G}^2 \over 2 G^2} - {{\ddot G} \over G}
    \right) 
    \Bigg]{\tilde v}
    -2(3a^2 H {\dot {\tilde v}}+{\dot \Delta}+3{\dot \Psi})\,,\label{Eq:Ecop}\\
    \calE^{\rm eu} &=& 
    \Phi + a^2(2H {\tilde v} + {\dot {\tilde v}})
    \,,\label{Eq:Eeup}
\ea
where we defined
\ba
    \calA &=& M_{\rm GW}^2 -{3(\rho_g + p_g) \over \mpl^2} -6 m^2 \xi^4 \Big(
    U_{[Z][Z]}+4\xi^2 U_{[Z][Z^2]}+ 6\xi^4 U_{[Z][Z^3]} + 4 \xi^4 U_{[Z^2][Z^2]}+ 12 \xi^6 U_{[Z^2][Z^3]}+ 9\xi^8 U_{[Z^3][Z^3]} 
    \Big)\,, \notag\\
    \\
    \calC_B &=& 
    2 {\dddot G} - {6 {\dot G}{\ddot G} \over G} 
    + 8 H {\ddot G} + {3{\dot G}^3 \over G^2} 
    -{12 H {\dot G}^2 \over G} + 2 \calA H + 
    {\rho + \rho_g -3p_g \over \mpl^2} {{\dot G} \over G}
    + {2{\dot p}_g \over \mpl^2}
    \,,\\
    \calC_{\dot S} &=& 
    3 {\dddot G} - {21 {\dot G}{\ddot G} \over 2G} 
    + 18 H {\ddot G} + {6{\dot G}^3 \over G^2} 
    -{24 H {\dot G}^2 \over G} + 3 \calA H + 
    12 H^2 {\dot G}
    +{\rho + \rho_g -6p_g \over \mpl^2} {{\dot G} \over G}
    + {3{\dot p}_g \over \mpl^2}\,.
\ea

As we saw in Section~\ref{sec:QS}, in the minimal coupling model, we can construct the master equations for $S$ and $\Delta$ from the linear equations of motion. At subhorizon scales, their quasistatic limit exactly follows the numerical solution. Similarly, in the model with nonminimal coupling, we can construct the master equations for $S$ and $\Delta$, but they are too complicated to show in this paper. One can find that the leading term in the quasistatic limit leads to $\Delta = 0$. However, this may be a signal that the quasistatic limit does not work properly. 
As we will see in Appendix~\ref{sec:nonminimal_coupling_model}, a simple nonminimal coupling model unfortunately suffers from perturbative instabilities, and a further investigation of the quasistatic limit for stable nonminimal coupling models will be performed in the future.

\section{Nonminimal coupling model : $G(X) \neq 1$ }
\label{sec:nonminimal_coupling_model}

In this appendix, we provide a supplementary discussion for background stabilities in the case of a non-minimal coupling model.

\subsection{Concrete model and stabilities}
We will consider the concrete model of projected massive gravity with the following nonminimal coupling function,
\begin{align}
  G(X) = 1 + g m^2 X\,,
  \label{eq:nonminimalcoupling}
\end{align}
and the mass potential 
\begin{align}
U(X,[Z],[Z^2],[Z^3])=& (a_1 + a_2 X)[Z] + (b_1 + b_2 X) [Z]^2 + (c_1 + c_2 X) [Z^2] \notag\\
  &  + (d_1 + d_2 X) [Z]^3+ (e_1 + e_2 X) [Z][Z^2]+ (f_1 + f_2 X) [Z^3] \,,
\end{align}
where $g, a_1 a_2,b_1,b_2,c_1,c_2,d_1,d_2,e_1,e_2,f_1,$ and $f_2$ are dimensionless constant parameters. In this model, the St\"{u}ckelberg equation \eqref{StukelbergEqn} becomes
\begin{align}
    \xi  (H \xi+\dot{\xi} )F(\xi^2)=0\,,
    \label{eq:StukelbergEqn_nonminimal}
\end{align}
where $F(\xi^2)$ is defined by
\begin{align}
    F(\chi)=&-5  \left[g \mu^4 a^5 (9 d_2+3e_2+f_2)\right] \chi^4+ \left[4 \mu^2 \Omega_\kappa  a^3 (9 d_1 g+9 d_2+3 e_1 g+3 e_2+g f_1+f_2)-3 g \mu^4 a^5 (3 b_2+c_2)\right]\chi^3 \notag\\
    &+ \left[-a_2 g \mu^4 a^5+\mu^2 \Omega_\kappa  a^3 (6 b_1 g+9 b_2+2 c_1 g+3 c_2)-3 \Omega_\kappa^2 a (9 d_1+3 e_1+f_1)\right]\chi^2 \notag\\
    &+\left[2 a_2 \mu^2 \Omega_\kappa  a^3-2 \Omega_\kappa^2 a (3 b_1+c_1)\right]\chi+ \Omega_\kappa\left(g \Omega_m-a_1 \Omega_\kappa a\right)\,,
\end{align}
and $\chi := \xi^2$. Assuming $\xi$ is nonzero, the solutions of \eqref{eq:StukelbergEqn_nonminimal} are given by $\dot{\xi}+H\xi=0$ and $F(\chi)=0$. In the former case, $\rho_g$ behaves as the sum of the spacial curvature and the radiation, then we are not interested in this case. For the later case, since $F(\chi)$ is a quartic function of $\chi$, it is difficult to analyze generally. For simplicity, we assume the following relations of parameters. 
\begin{align}
    f_1 = -3 (3 d_1 + e_1 )\,,\quad f_2 = -3 (3 d_2 + e_2 )\,,\quad c_2=-3 b_2\,.
\end{align}
In this case, $F(\chi)=0$ becomes a quadratic equation and its solutions are
\begin{align}
  \chi=\chi_{\pm}:=\frac{B\Omega_\kappa^2-a_2 \Omega_\kappa a^2 \pm \sqrt{R}}{g\mu^2 a^2 (2B\Omega_\kappa-a_2 a^2)}\,,
  \label{sol_stackelberg}
\end{align}
with 
\begin{align}
  R(a) := \Omega_\kappa \left[ B^2 \Omega_\kappa^3 + a (a_2 a^2-2B\Omega_m) \left( (a_2-a_1 g)\Omega_\kappa a+g^2 \Omega_m \right)\right]\,,
  \qquad
  \mu^2 B := c_1 + 3b_1\,.
\end{align}

To get the real $\xi$, we impose the positivity of $R$ and $\chi_{\pm}$ at least in the limit of $a\to 0$ and $a\to \infty$. In the case $a \to \infty$, the solution $\chi_{\pm}$ asymptotically behaves as
\begin{align}
    \chi_{\pm} &\approx \frac{ \Omega_\kappa}{a_2g a^2}\left( a_2\mp \sqrt{a_2 (a_2-a_1 g)} \right)\,,
\end{align}
and then, $a_2$ should be positive.

In the following, we will assume $a_2 \neq 0 \land B \neq 0$. In this case, we can show that it is not possible to avoid ghost instability. \footnote{
Even in the case of $B\neq0 \land a_2 \neq 0$, $B=0 \land a_2 \neq 0$ and $B\neq0 \land a_2=0$, the ghost instability is not inevitable. In the case of $B=0 \land a_2=0 \land a_1\neq0$, the expanding solution does not exist, but the Minkowski solution exist. }

\subsection{No ghost condition}

Substituting the solutions \eqref{sol_stackelberg} into \eqref{eq:nonminimalcoupling}, we obtain the conformal factor $G=G_{\pm}$ for each solution with
\begin{align}
  G_\pm := \frac{B\Omega_\kappa^2 \mp \sqrt{R}}{\Omega_\kappa(2B\Omega_\kappa-a_2 a^2) }\,.
  \label{conformal_factor}
\end{align}
At large $a$, $G_{\pm}$ asymptotically approaches
\begin{align}
    \lim_{a \to \infty} G_{\pm}=\pm \frac{\sqrt{a_2 (a_2-a_1 g)}}{a_2}\,.
\end{align}
To avoid the ghost instability for the tensor gravitons, we impose the positivity of $G_\pm$. In the limit of $a \to \infty$, it translates $a_2 > 0 ~\land~ a_2-a_1 g \geq 0$ for $\chi_+$, and $a_2 < 0 ~\land~ a_2-a_1 g \leq 0$ for $\chi_-$. Note that even in the case of $a_2 B < 0$, either \eqref{sol_stackelberg} or \eqref{conformal_factor} is not singular at $a=\sqrt{2B\Omega_\kappa/a_2}$ if these conditions are held.

In the small $a$ limit, both $G_{\pm}$ approach zero if $B a_2 < 0$, and the derivative of $G_{\pm}$ with respect to $a$ approach
\begin{align}
  \frac{dG_{\pm}}{da} \approx  \frac{g^2\Omega_m}{2B\Omega_\kappa^2}\,.
\end{align}
Then, $a_2<0 \land B>0$ is not allowed because $G_\pm$ is negative at an early time. Therefore, the parameter regions without ghost for tensor gravitons are
\begin{enumerate}
    \item $B\neq 0 ~\land~ a_2 > 0 ~\land~ a_2-a_1 g \geq 0 \,, \mathrm{~for~}\chi_+$\,,
    \item $B > 0 ~\land~ a_2 < 0 ~\land~ a_2-a_1 g \leq 0\,,\mathrm{~for~}\chi_-$\,.
\end{enumerate}

To avoid the ghost instability for the scalar graviton, we also impose 
\begin{align}
    \rho_g+p_g = -\frac{H_0^2 \mpl^2}{g^2 \Omega_\kappa a^4 (2B \Omega_\kappa-a_2 a^2)}\left[\pm\sqrt{R}+\Omega_\kappa (B\Omega_\kappa-a_2 a^2)\right]\left[\pm\sqrt{R}+\Omega_\kappa (B\Omega_\kappa-(a_2-a_1 g) a^2)\right] \geq 0
    \label{NEC_sol}
\end{align}
In the small $a$ limit, $\rho_g+p_g$ asymptotically behaves as 
\begin{align}
    \frac{\rho_g + p_g}{H_0^2\mpl^2} &\approx -\frac{2 }{g^2 a^4} < 0\,,\mathrm{~for~} \chi_+\mathrm{~with~}B > 0\,,
    \\
    \frac{\rho_g + p_g}{H_0^2\mpl^2} &\approx -\frac{g^2  \Omega_m^2}{2B\Omega_\kappa^2 a^2}>0\,,\mathrm{~for~} \chi_+\mathrm{~with~}B < 0\,,
    \\
    \frac{\rho_g + p_g}{H_0^2\mpl^2} &\approx -\frac{g^2  \Omega_m^2}{2B\Omega_\kappa^2 a^2}<0\,,\mathrm{~for~} \chi_-\mathrm{~with~}B > 0\,.
\end{align}
At the early time, the first and third cases violate the ghost free condition, and only the second case survives the ghost-free condition.
In the large $a$ limit, $\rho_g+p_g$ behaves as
\begin{align}
    &\frac{\rho_g+p_g}{H_0^2\mpl^2}\approx\frac{ \Omega_\kappa  }{a_2 g^2 a^2}\left(\sqrt{a_2 (a_2-a_1 g)}-a_2\right) \left(\sqrt{a_2 (a_2-a_1 g)}+a_1 g-a_2\right)<0\,,
\end{align}
for $\chi_+$ with $a_2 >0 \land a_2-a_1 g\geq 0$. Hence, this model can never avoid ghost problem.

Note that the model with a nonminimal coupling
\begin{align}
    G(X)=(1+gm^2X)^2\,,
\end{align}
can avoid the ghost and gradient instabilities at least in the limit of $a\to\infty$ and 0 as in the similar manner done in above. 
However, we numerically checked that even this model suffers from at least one of ghost, gradient, and tachyonic instabilities at $a=1$.


\bibliography{ref}

\end{document}